\documentclass[aps,prr,reprint, amsmath, amssymb,superscriptaddress,longbibliography]{revtex4-1}
 
\usepackage{bm}
\usepackage[retainorgcmds]{IEEEtrantools}
\usepackage{graphicx}
\usepackage{mathrsfs}
\usepackage{amsmath}
\usepackage{amsfonts}
\usepackage{amssymb}
\usepackage{color}
\usepackage{times,txfonts}
\usepackage{nicefrac}
\usepackage{ragged2e}
\usepackage{tikz}
\usepackage{bbold}

\usepackage[colorlinks=true,linkcolor=blue,urlcolor=blue,citecolor=blue,pdfusetitle]{hyperref}

\usepackage{blindtext}

\newcommand{\tr}{\mathrm{tr}}

\newcommand{\ket}[1]{\ensuremath{|#1\rangle}}
\newcommand{\bra}[1]{\langle#1|}
\newcommand{\braket}[2]{\langle#1|#2\rangle}
\newcommand{\ketbra}[2]{|#1\rangle\langle#2|}

\definecolor{cadmiumgreen}{HTML}{097969}

\begin{document}

\title{Entropy of the quantum work distribution}
\date{\today}
\date{\today}
\author{Anthony Kiely}
\email{anthony.kiely@ucd.ie}
\affiliation{School of Physics, University College Dublin, Belfield, Dublin 4, Ireland}
\affiliation{Centre for Quantum Engineering, Science, and Technology,
University College Dublin, Belfield, Dublin 4, Ireland}
\author{Eoin O'Connor}
\affiliation{School of Physics, University College Dublin, Belfield, Dublin 4, Ireland}
\affiliation{Centre for Quantum Engineering, Science, and Technology,
University College Dublin, Belfield, Dublin 4, Ireland}
\author{Thom\'as Fogarty}
\affiliation{Quantum Systems Unit, Okinawa Institute of Science and Technology Graduate University, Onna, Okinawa 904-0495, Japan}
\author{Gabriel T. Landi}
\affiliation{Instituto de F\'isica da Universidade de S\~ao Paulo,  05314-970 S\~ao Paulo, Brazil.}
\affiliation{School of Physics, Trinity College Dublin, College Green, Dublin 2, Ireland}
\author{Steve Campbell}
\affiliation{School of Physics, University College Dublin, Belfield, Dublin 4, Ireland}
\affiliation{Centre for Quantum Engineering, Science, and Technology,
University College Dublin, Belfield, Dublin 4, Ireland}

\begin{abstract}
The statistics of work done on a quantum system can be quantified by the two-point measurement scheme. 
We show how the Shannon entropy of the work distribution admits a general upper bound depending on the initial diagonal entropy, and a purely quantum term associated to the relative entropy of coherence. 
We demonstrate that this approach captures strong signatures of the underlying physics in a diverse range of settings. In particular, we carry out a detailed study of the Aubry-Andr\'e-Harper model and show that the entropy of the work distribution conveys very clearly the physics of the localization transition, which is not apparent from the statistical moments.
\end{abstract}

\maketitle{}

{\bf \emph{Introduction.--}}Work in a quantum mechanical setting has proven to be a difficult concept to define \cite{campisi2011}, with several approaches developed~\cite{Alicki1979,Kosloff2013,Frenzel2014,Dahlsten2017,Deffner2016,Alipour2022}. Among them the two-point measurement (TPM) approach \cite{Talkner2007} has received significant attention: it recovers important results from stochastic thermodynamics~\cite{Deffner2008,Goold2016}, can be measured experimentally \cite{Batalhao2014,DeChiara2015} and naturally connects with other areas, such as, out-of-time-order correlators~\cite{Campisi2017}, information scrambling~\cite{ChenuSciRep,ChenuQuantum}, Kibble-Zurek scaling~\cite{delCampoPRL,Fei2020}, and many-body physics~\cite{Fusco2014}.
Often the focus is on cumulants of work (in particular the mean and variance) rather than the full distribution. While in several contexts this is warranted, particularly when the underlying distribution tends to a Gaussian~\cite{Zawadzki2022}, several recent works have highlighted that studying the full distribution can reveal non-trivial features of the dynamics that, while perhaps present in the statistical cumulants, are nevertheless obfuscated~\cite{Zawadzki2020,Zawadzki2022}. 

Recently, it has been shown that coherence plays a subtle role in establishing a proper thermodynamic framework~\cite{JuanDelgado2021,Diaz2020,Korzekwa2016,goold2018}. Indeed, quantum coherences present a viable source of useful work~\cite{Klatzow2019} and, as such, there is an intrinsic thermodynamic cost associated with their creation~\cite{Huber2015,Misra2016}. However, while potentially useful, the presence or creation of coherence when a system is driven out-of-equilibrium can lead to significant fluctuations~\cite{Miller2020}. A more careful analysis of such non-equilibrium dynamics reveals that one can identify uniquely quantum aspects in the thermodynamics of quantum systems, in particular by splitting the irreversible work into distinct coherent and incoherent contributions~\cite{Santos2019,Francica2019}. While these, and related studies~\cite{Fusco2014}, have focused on the moments, it is intuitive that the full distribution should encapsulate and extend these insights.

We rigorously demonstrate the veracity of this intuition through the entropy $H_W$ of the work distribution, which serves as a measure of its underlying complexity. This measure has been applied to the distribution of entropy production \cite{Salazar2021}. We derive a general and saturable bound on $H_W$, that consists of two distinct contributions: one which stems from the diagonal ensemble and, in suitable limits, corresponds simply to the Gibbs equilibrium entropy; and a second term which is purely quantum in nature, related to the coherence established by the driving protocol, and given by the relative entropy of coherence. 
We first illustrate the utility of our results in the  Landau-Zener model which reveals that the entropy of the distribution succinctly captures the salient features of the model around the avoided crossing, features which are completely absent in the moments. 
We then carry out a detailed analysis of work fluctuations in the Aubry-Andr\'e-Harper (AAH) model, a paradigmatic model for studying localization.
We show that $H_W$ is related to a modified inverse participation ratio, and provides a remarkably sensitive indicator of the localization transition.

{\bf \emph{Entropy of the Work Distribution.--}}We consider a system,  prepared in a generic state $\rho$, and with initial Hamiltonian $\mathcal{H}_i=\sum_n E_n^i |n_i\rangle\langle n_i|$, that is driven according to a work protocol, which changes the state to $\rho' \!=\! U \rho U^\dagger$. The unitary $U$ depends on the details of the protocol and its duration. The Hamiltonian at the end of the process is $\mathcal{H}_f \!=\! \sum_m E_m^f |m_f\rangle\langle m_f|$. The TPM consists of measuring in the bases of $\mathcal{H}_i$ and $\mathcal{H}_f$, before and after the unitary~\cite{Talkner2007}. The probability that a certain amount of work, $W$, is injected or extracted is given by
\begin{equation}\label{PW}
    P(W) = \sum\limits_{n,m} ~p_n p_{m|n} \delta_{W,E_m^f - E_n^i}\;,
\end{equation}
where $p_n \!=\! \langle n_i |\rho|n_i\rangle$ is the initial state distribution and $p_{m|n}\!=\!|\langle m_f | U| n_i \rangle|^2$ are the transition probabilities. The support of $P(W)$ corresponds to all possible \emph{Bohr (transition) frequencies} $E_m^f - E_n^i$ between the initial and final energy levels. 
We assume these form a discrete (possibly infinite) set. Note how Eq.~\eqref{PW} they are collected in different pairs $(n,m)$ which give rise to the same value of $W$. 

The work distribution can be very complex, so one often  focuses on summary statistics, such as the  moments $\langle W^n \rangle \!=\! \sum_W W^n P(W)$, or cumulants. Here, we shift focus to another summary statistic; namely, the  entropy of $P(W)$~\footnote{The work distribution~\eqref{PW} is usually defined with a Dirac delta. As far as the entropy is concerned, however, the discreteness of the support is important when dealing with the entropy, which is why we defined it here with a Kronecker delta instead.}:
\begin{equation}\label{HW}
    H_W = - \sum_W P(W) \ln P(W),
\end{equation}
which characterizes the complexity of $P(W)$. It is zero when the work is deterministic and can range up to  $\ln N^2$ when $P(W)$ is uniform.

$H_W$ is in general different from 
\begin{equation}\label{Hu}
    H_{\rm u} = -\sum_{n,m} p_n p_{m|n}\ln{p_n p_{m|n}},
\end{equation}
which is the entropy of the \emph{uncollected} distribution $p_n p_{m|n}$. We first quantify the relation between $H_W$ and $H_{\rm u}$. 
Let $\gamma_{\rm max}$ denote the maximal degeneracy of the Bohr frequencies ($\gamma_{\rm max} \geq g_i g_f$, where $g_{i(f)}$ are the degeneracies of $\mathcal{H}_{i(f)}$).
Then~\cite{SupMat}
\begin{equation}\label{bounds_W_u}
    H_{\rm u} - \ln \gamma_{\rm max} \leq H_W \leq H_{\rm u},
\end{equation}
with equality if the values of work are all non-degenerate. 
We now show that $H_{\rm u}$ directly quantifies the degree of quantum coherence generated in the process. The relative entropy of coherence (REC)~\cite{Baumgratz2014} of a state $\sigma$ in the basis $\ket{m_f'}\!=\!U^\dagger \ket{m_f}$ is
\begin{eqnarray}\label{relative_entropy_coherence}
    C(\sigma)=S(D_f(\sigma))-S(\sigma) \geq 0,
\end{eqnarray}
where $S(\sigma)\!=\!-\tr\,\sigma \ln \sigma$ is the von-Neumann entropy and $D_f(\sigma)\!=\!\sum_m \bra{m_f'}\sigma \ket{m_f'} \ketbra{m_f'}{m_f'}$ is the full dephasing operation in the basis $\ket{m_f'}$. 
It follows that $-\sum_m p_{m|n} \ln p_{m|n} \!=\! \mathcal{C}\big(\ketbra{n_i}{n_i}\big)$, so Eq.~\eqref{Hu} can be written as 
\begin{equation}\label{Hu2}
    H_{\rm u} = S(\bar{\rho}) + \sum_n p_n \mathcal{C}\big(\ketbra{n_i}{n_i}\big),
\end{equation}
where $\bar{\rho} \!=\! \sum_n \bra{n_i}\rho \ket{n_i} \ketbra{n_i}{n_i}$ is the initial state dephased in the basis of $\mathcal{H}_i$. 

Equation~\eqref{Hu2} summarizes the rich physics behind the entropy of the work distribution. The first term is the entropy of the initial outcomes $p_n$ of the TPM, i.e. the entropy of the so-called diagonal ensemble~\cite{Pietracaprina2017,Cakan2021,Goold2015,Mzaouali2021,Wang2021}. If $[\rho,\mathcal{H}_i]\!=\!0$, it reduces to the von Neumann entropy of $\rho$ and if $\rho \!=\! e^{-\beta \mathcal{H}_i}/Z_i$ is a thermal state, it reduces to the Gibbs thermal entropy. If $\rho \!=\! |k_i\rangle\langle k_i|$ is any eigenstate of $\mathcal{H}_i$, $S(\bar{\rho})$  vanishes and Eq.~\eqref{Hu2} reduces to 
$H_{\rm u} \!=\! \mathcal{C}\big(\ketbra{k_i}{k_i}\big)$.
The second term in Eq.~\eqref{Hu2} establishes that the relevant coherences are those of each $|n_i\rangle$ in the eigenbasis $|m_f'\rangle$. Therefore, this term contains information on both the dynamics (work protocol) and of how $\mathcal{H}_f$ differs from $\mathcal{H}_i$. The process is  incoherent if $p_{m|n} \!=\! |\langle m_f |U |n_i\rangle|^2 \!=\! \delta_{m,n}$, which occurs when $[\mathcal{H}_i, U^\dagger \mathcal{H}_f U]\!=\!0$.  In this case, Eq.~\eqref{Hu2}  reduces to $H_{\rm u} \!=\! S(\bar{\rho})$. 

We can take this a step further. Using the concavity of the von Neumann entropy, we can write $\sum_n p_n \mathcal{C}\big(\ketbra{n_i}{n_i}\big) \!\leq\! S\big(D_f(\bar{\rho})\big) \!=\! \mathcal{C}(\bar{\rho}) + S(\bar{\rho})$, which leads to
\begin{equation}\label{bound_u_2}
    H_{\rm u} \leq 2 S(\bar{\rho}) + \mathcal{C}(\bar{\rho}).
\end{equation}
The tightness of this bound is related to the purity of $\bar{\rho}$, being saturated when $\rho$ is an eigenstate of $\mathcal{H}_i$ or for thermal states in the zero temperature limit.

Combining Eqs.~\eqref{bounds_W_u},~\eqref{Hu2} and~\eqref{bound_u_2}, we arrive at our main result: the entropy of the work distribution is bounded as 
\begin{equation}\label{bound_W_ultimate}
    H_W \leq S(\bar{\rho}) + \sum_n p_n \mathcal{C}\big(\ketbra{n_i}{n_i}\big) \leq 2 S(\bar{\rho}) + \mathcal{C}(\bar{\rho}).
\end{equation}
The first inequality is often quite tight, and relates $H_W$ to the coherences of each individual transition $\mathcal{C}\big(\ketbra{n_i}{n_i}\big)$. The second inequality bounds $H_W$ to the full REC of $\bar{\rho}$ and its tightness is related to the purity of $\bar{\rho}$. Eq.~\eqref{bound_W_ultimate} also allows us to estimate the dependence of $H_W$ with temperature $T$, in the case of an initial thermal state. Both $S(\bar{\rho})$ and the $p_n$ depend on $T$. However, by convexity
\begin{equation}\label{temperature_dependence}
    H_W \leq S(\bar{\rho}) + \mathcal{C}_{\rm max},
\end{equation}
where $\mathcal{C}_{\rm max}\! =\! \max\limits_n \mathcal{C}\big(\ketbra{n_i}{n_i}\big)$. 
The last term is now $T$ independent, pushing the temperature dependence solely to the Gibbs thermal entropy.
We next turn to the study of $H_W$ in different models, and show that it conveys crucial information about the work statistics.

{\bf \emph{Landau-Zener model.--}}Consider a qubit with $\mathcal{H}_{\rm LZ}(\omega)=\hbar \Delta\sigma_x+\hbar \omega \sigma_z$ where $\sigma_i$ are the Pauli matrices. This model has an avoided crossing at $\omega_c\equiv 0$, with minimal energy gap $\Delta\!>\!0$. The eigenenergies are $E_0\!=\!-\sqrt{\omega^2+\Delta^2} \hbar$ and $E_1\!=\!\sqrt{\omega^2+\Delta^2} \hbar$. We assume the system starts in a thermal state at inverse temperature $\beta$, and consider a sudden quench ($U\!=\mathbb{1}$) from $\mathcal{H}_i\!=\!\mathcal{H}_{\rm LZ}(\omega_i)$, with $\omega_i<0$, to $\mathcal{H}_f\!=\!\mathcal{H}_{\rm LZ}(\omega_f)$. There are four allowed values of $W$, given by $E_{0(1)}(\omega_f) - E_{0(1)}(\omega_i)$.
For $\omega_f \!\neq\! \pm \omega_i$ and fixed $\Delta$, these will always be non-degenerate and thus $H_W\!\equiv \!H_{\rm u}$. 
\begin{figure}[t]
\begin{center}
\includegraphics[angle=0,width=0.98\linewidth]{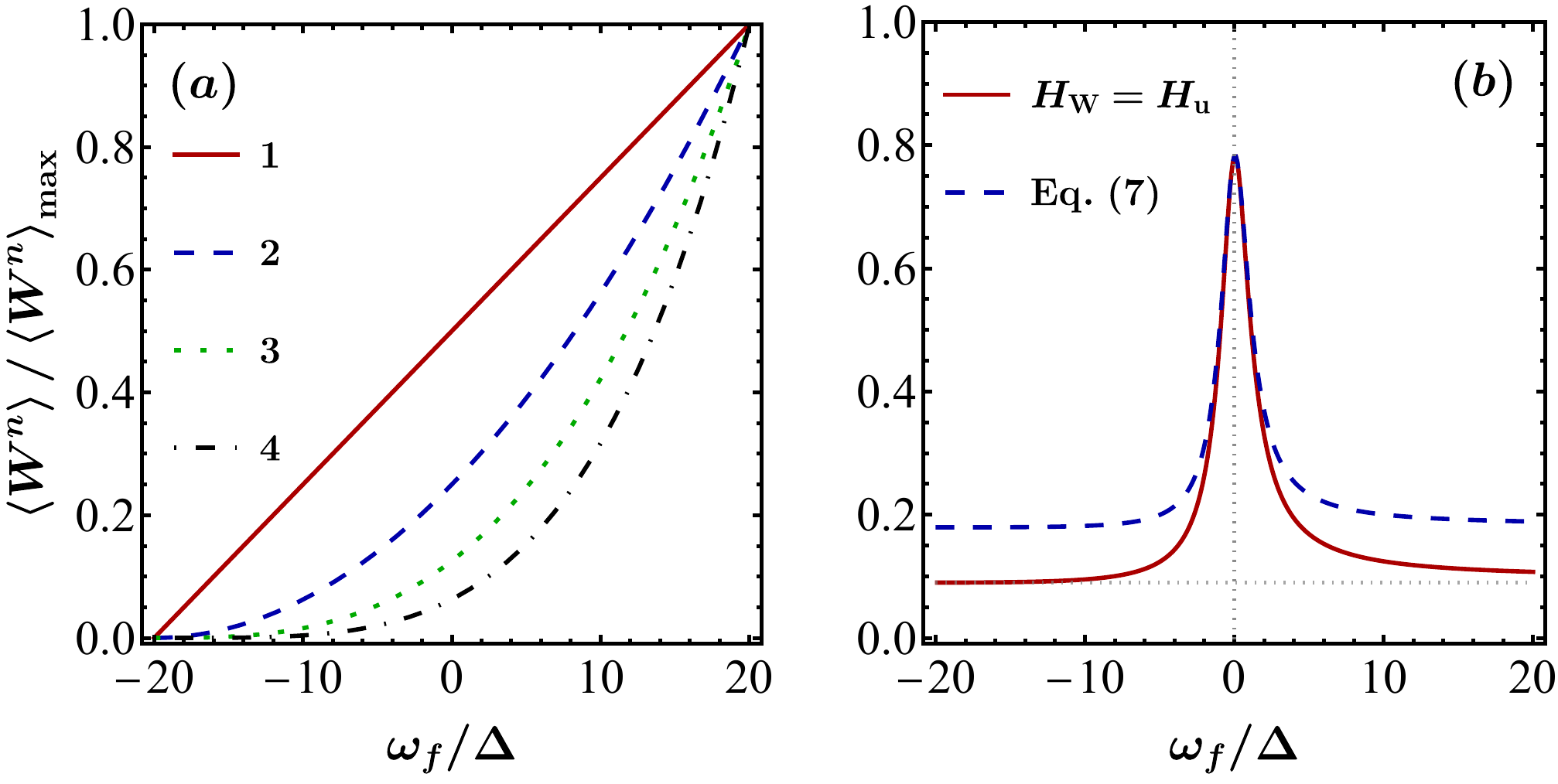} 
\end{center}
\caption{Work fluctuations in the Landau-Zener model under a sudden quench. (a) First four moments $\langle W^n\rangle$ of $P(W)$ as a function of $\omega_f/\Delta$ (normalized by their maximum value, at $\omega_f =\Delta$).
(b) Entropy of the work distribution, Eq.~\eqref{HW} (red-solid), and the corresponding bound~\eqref{bound_u_2} (blue-dashed).
Parameters: $\beta = 0.1(\hbar\Delta)^{-1}$ and $\omega_i=-20 \Delta$.
}
\label{fig_LZ}
\end{figure}

Figure~\ref{fig_LZ}(a) shows the first four moments $\langle W^n\rangle$ of $P(W)$, as a function of $\omega_f/\Delta$, while Fig.~\ref{fig_LZ}(b) shows $H_W$. Clearly the moments show no obvious evidence of the avoided crossing at $\omega_f \!=\! \omega_c$ (the same is true for the cumulants). The entropy $H_W$, on the other hand, portrays an entirely different picture.
The first term in~\eqref{Hu2} yields a constant base value, as it depends only on the initial condition. 
The second term, on the other hand, presents a  peak at $\omega_f \!=\! \omega_c$. By probing $H_W$ we can therefore highlight the avoided crossing, which is the most important feature of the Landau-Zener model, and which is masked in the moments. 
In Fig.~\ref{fig_LZ}(b) we also plot the bound~\eqref{bound_u_2}, which becomes tightest around $\omega_f \!=\! 0$. This reflects the coherence, which is largest at the avoided crossing. 

\begin{figure*}
    \centering
    \includegraphics[width=\textwidth]{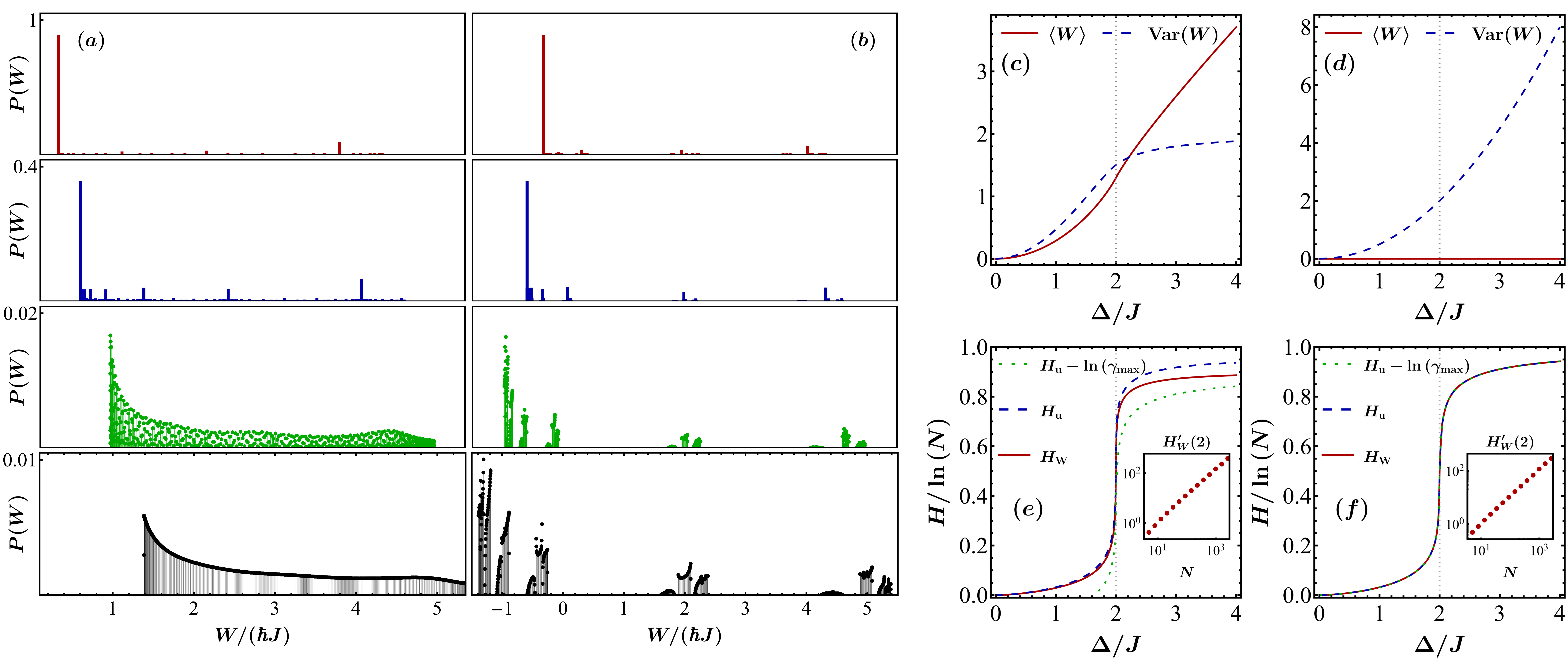}
    \caption{Work statistics of the AAH model~\eqref{H_AAH}.
    (a) $P(W)$ for the $\Delta\to 0$ protocol, for $\Delta/J = 1.5,2,2.5$ and $3$.
    (b) Similar, but for $0\to\Delta$.
    (c,d) Corresponding mean and variance~vs.~$\Delta/J$, for the two protocols.
    In (d) $\langle W \rangle \equiv 0$~\cite{SupMat}.
    (e,f) $H_W$~vs.~$\Delta/J$ [Eq.~\eqref{HW}] for the two protocols along with the upper and lower bounds derived in Eq.~\eqref{Hu2}. (Inset): $dH_W/d\Delta \big|_{\Delta = 2J}$ as a function of a Fibonacci number $N$, showing that in the thermodynamic limit $H_W$ will change discontinuously at $\Delta/J=2$.
    In all simulations the system starts in the ground-state, $N =F_{16} =987$ and $\eta = 1.2$,
    except in the insets of (e,f), which were averaged over 50 values of $\eta$. 
    }
    \label{fig:AAH_panel}
\end{figure*}

{\bf \emph{AAH~model.--}}
We next turn to a highly  non-trivial application of our results. We consider a single particle in a lattice with $N$ sites, labeled by states $|i\rangle$. 
The Hamiltonian is~\cite{Aubry1980,Harper1955,DomnguezCastro2019}
\begin{equation}\label{H_AAH}
    \mathcal{H}_{\rm AAH}(\Delta) = \hbar \sum_{i=1}^N \Bigg[\Delta\cos\left(2 \pi \gamma i + \eta \right) |i\rangle\langle i| - J \big(|i\rangle\langle i+1| + |i+1\rangle\langle i|\big)\Bigg],
\end{equation}
with periodic boundary conditions. The first term denotes the on-site potentials, with overall magnitude $\Delta$, phase $\eta$, and modulation $\gamma$. 
Following~\cite{Lye2007,Tanzi2013,DomnguezCastro2019}, we choose the lattice size $N$ to be a Fibonacci number, $F_n$ and $\gamma \!=\! F_{n-1}/F_n$ to be a rational approximation to the inverse golden ratio~\footnote{This ensures that while the periodic boundary conditions are fulfilled, the potential is aperiodic within the lattice as in experimental realizations in optical lattices.}.

The AAH model undergoes a localization transition at $\Delta \!=\! 2J$. 
For $\Delta\!<\!2J$ all eigenvectors are delocalized in space,
while for $\Delta \!>\! 2J$, they become localized around specific sites in the lattice. 
We focus on the work distribution associated with turning the quasiperiodic potential off/on, i.e. in going from $\mathcal{H}_{\rm AAH}(\Delta)\to \mathcal{H}_{\rm AAH}(0)$, and vice-versa. We refer to these as $\Delta\!\to\!0$ and $0\!\to\!\Delta$, respectively, and in what follows we focus on sudden quenches ($U=\mathbb{1}$).

Fig.~\ref{fig:AAH_panel}(a,b) shows the work distribution~\eqref{PW} for the two protocols, assuming the system starts in the ground-state. The bandwidth of the distribution is discussed in~\cite{SupMat}. For $\Delta\! \to\! 0$, $W>0$, while for $0\to\Delta$, $W \lessgtr 0$. Thus, work can be extracted by turning the potential on, but not by turning it off. The overall behavior of $P(W)$ clearly reflects the localization transition at $\Delta\!=\!2J$. For both protocols, quenches that keep the system in the delocalized phase, i.e. $\Delta\!<\!2J$ (corresponding to the first two upper panels of Fig.~\ref{fig:AAH_panel}(a,b)), result in a $P(W)$ with small support, and mostly concentrated around a minimum work value. In this regime, the work cost of turning the potentials on or off is overall small and fluctuates very little. This is also evidenced in Fig.~\ref{fig:AAH_panel}(c,d), which plots the mean and variance of $W$, for the two protocols. Conversely, when $\Delta/J\!>\!2$ the support of $P(W)$ increases significantly. For $\Delta\!\to\! 0$ (Fig.~\ref{fig:AAH_panel}(a)) the distribution reflects the smooth energy spectrum, while for $0\!\to\!\Delta$ [Fig.~\ref{fig:AAH_panel}(b)] it is very irregular due to the fractal nature of the localized spectrum. 

\begin{figure}[b]
\begin{center}
\includegraphics[angle=0,width=0.98\linewidth]{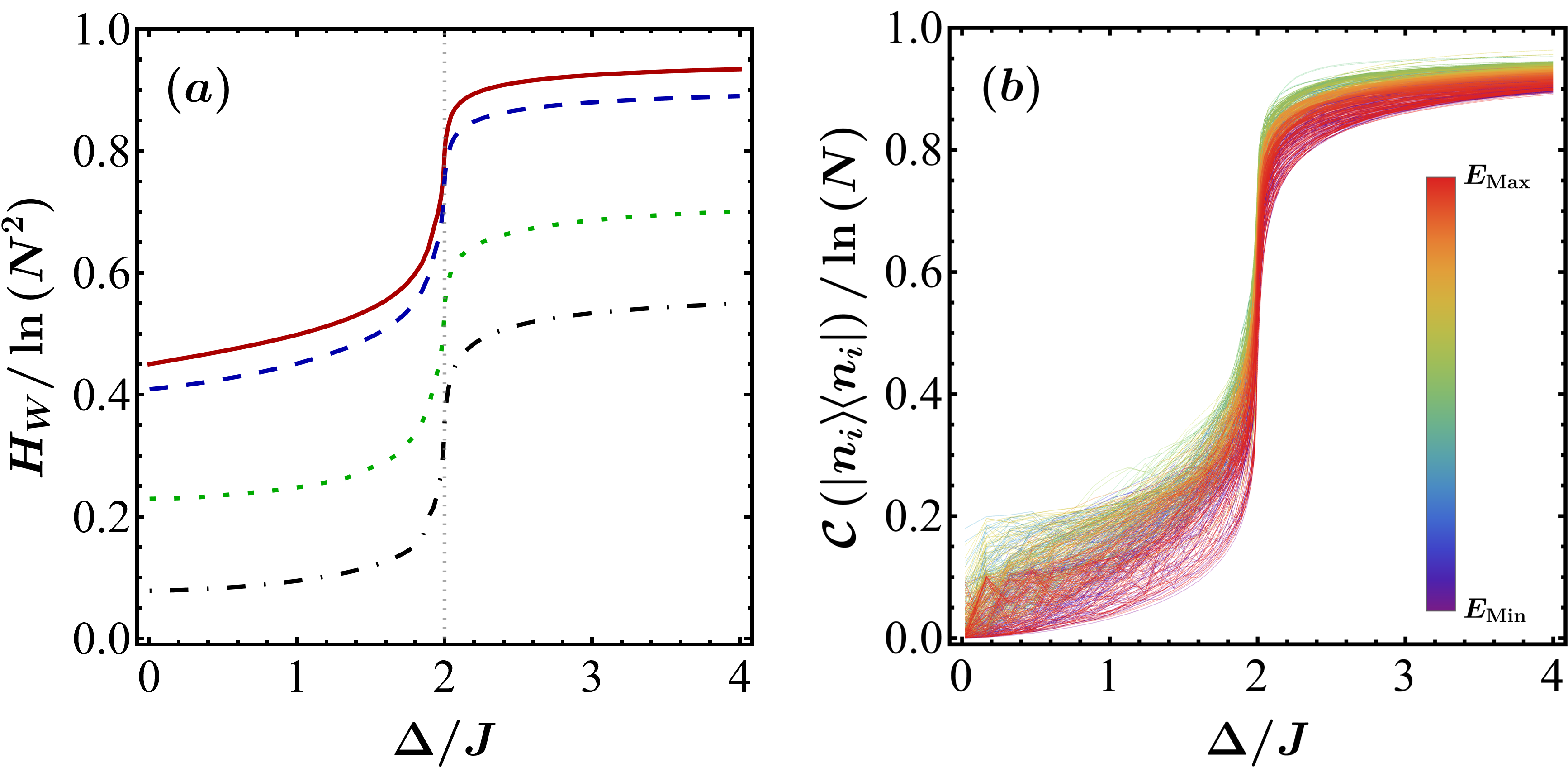} 
\end{center}
\caption{The entropy of $P(W)$ for (a) initial thermal states with temperatures $J \beta=\{10^{-2},10^0,10^2,10^4\}$(red [top], blue, green, black [bottom]) and (b) every eigenstate of the initial Hamiltonian, $\mathcal{H}_{\rm AAH}(0)$. These are all for the $0\to \Delta$ case but the $\Delta \to 0$ case is very similar. The choice of phase and system size are as in Fig.~\ref{fig:AAH_panel}.
}
\label{fig:AAH_temp}
\end{figure}

$H_W$ is plotted in Figs.~\ref{fig:AAH_panel}(e,f). 
It shows a jump at $\Delta/J \!=\! 2$, the sharpness of which depends on the lattice size $N$. To illustrate this, the insets of Figs.~\ref{fig:AAH_panel}(e,f) show the slope $H_W'(2)\!:=\! dH_W/d\Delta$, evaluated at $\Delta \!=\! 2J$, for different sizes $N$. A fit of the data reveals the relation, $H_W'(2)\!\propto\! \sqrt{N}$, which implies that, in the thermodynamic limit, $H_W$ will change discontinuously at the localization transition. The entropy therefore succinctly captures the criticality of the AAH model. 
The two bounds in Eq.~\eqref{bounds_W_u} are also shown in Figs.~\ref{fig:AAH_panel}(e,f).
For any $\Delta \neq0$, the spectrum of the AAH model is non-degenerate. 
This explains why for $\Delta \to 0$ (Fig.\ref{fig:AAH_panel}(e)) the curves differ from $H_W$, but for $0\to\Delta$ (Fig.\ref{fig:AAH_panel}(f)) they coincide: the former depends on the degeneracies of $\mathcal{H}_{\rm AAH}(0)$, leading to $\gamma_{\rm max} = 2$, while the latter does not since we start in the (non-degenerate) ground-state. 

$H_{\rm u}$ can be connected to a modified Inverse Participation Ratio (IPR), a widely used measure to characterize disordered systems.
The conventional IPR of a state $|\psi\rangle$ is defined as $\sum_{i} |\langle i| \psi\rangle|^4$, where $|i\rangle$ are the position states.
Instead, consider the quantity $\mathcal{I} := \sum_m p_{m|0}^2 = \sum_m |\langle m_f|0\rangle|^4$, where $0$ indexes the ground state. 
This is known as the inverse of the ``effective dimension"~\cite{Styliaris2019,Dooley2020}, and 
represents a type of IPR, where $|m_f\rangle$ replaces the position states $|i\rangle$ (they coincide if $\Delta_f \to \infty$).
Noticing that $-\ln \mathcal{I}$ is the R\'enyi-2 entropy of $p_{m|0}$, it then follows that
$ H_{\rm u} \geq -\ln \mathcal{I}$.
Hence, the physics of $H_W$ will reflect that of the modified IPR (the argument can also be extended to arbitrary initial states).




While Fig.~\ref{fig:AAH_panel} was concerned with the ground state, in the AAH model $H_W$ shows a qualitatively similar behavior at  finite temperatures [Fig.~\ref{fig:AAH_temp}(a)]. As the temperature increases $H_W$ tends to grow, but  maintains the same overall shape as a function of $\Delta$, and still exhibits strong signatures of the transition. 
This is due to the fact that in a localization transition \emph{all} eigenvectors undergo a sudden change.
As a consequence, all terms  $\mathcal{C}\big(|n_i\rangle\langle n_i|\big)$ in Eq.~\eqref{Hu2} will behave similarly, causing the bound~\eqref{temperature_dependence} to be fairly tight. 
We confirm this numerically in Fig.~\ref{fig:AAH_temp}(b), where we plot $\mathcal{C}\big(|n_i\rangle\langle n_i|\big)$ for all eigenvectors. 
We thus reach the conclusion that the monotonic vertical shift in $H_W$, observed in Fig.~\ref{fig:AAH_temp}(a), is essentially due to the Gibbs entropy $S(\bar{\rho})$. Our bounds therefore allow us to pinpoint different physical origins for different effects, namely thermal fluctuations and the localization transition. 


{\bf \emph{Conclusions.--}}We have demonstrated that the entropy of the quantum work distribution provides a useful tool in characterizing the non-equilibrium response of a quantum system. The entropy captures the complexity of the full distribution and we have shown that it is acutely sensitive to sudden changes in the system, such as avoided crossings and localization transitions. Our main result, Eq.~\eqref{bound_W_ultimate}, shows that $H_W$ can be understood as stemming from two distinct contributions, one given by the entropy of the initial state, dephased by the TPM, and a second term related to the coherences created by the work protocol. More specifically, what matters are the coherences of the initial eigenstates $|n_i\rangle$ in the basis $|m_f'\rangle = U^\dagger |m_f\rangle$. It therefore accounts not only for the change in Hamiltonian, from $\mathcal{H}_i \to \mathcal{H}_f$, but also for the entire work protocol, summarized by $U$. The contribution of quantum coherence to work has been explored in the past~\cite{Francica2019,Varizi2021,Scandi2019}, but only for initial thermal states, and with a focus on the average or the first few moments. 
Our results hold for any initial state, and also focus on a different quantity, thus being complementary. 
By means of examples, we have shown that the entropy is capable of conveying a richness of information that is not immediately visible in the moments. We therefore believe that it could serve as a powerful tool for characterizing work statistics away from equilibrium. Furthermore, these results could be extended to the density of states, which is a generalization of the work distribution \cite{Mzaouali2021}.

\emph{Acknowledgments --- }
The authors acknowledge fruitful discussions with Harry Miller, Gonzalo Manzano and Domingos Salazar. AK, EOC, and SC are supported by the Science Foundation Ireland Starting Investigator Research Grant ``SpeedDemon" No. 18/SIRG/5508. SC acknowledges the John Templeton Foundation (Grant ID 62422). This work was supported by the Okinawa Institute of Science and Technology Graduate University. TF acknowledges support under JSPS KAKENHI-21K13856. GTL  acknowledges the financial support of the Sa\~{o} Paulo Funding Agency FAPESP (Grant No. 2019/14072-0.) and the Brazilian funding agency CNPq (Grant No. INCT-IQ 246569/2014-0).

\bibliography{lib}

\begin{thebibliography}{50}%
\makeatletter
\providecommand \@ifxundefined [1]{%
 \@ifx{#1\undefined}
}%
\providecommand \@ifnum [1]{%
 \ifnum #1\expandafter \@firstoftwo
 \else \expandafter \@secondoftwo
 \fi
}%
\providecommand \@ifx [1]{%
 \ifx #1\expandafter \@firstoftwo
 \else \expandafter \@secondoftwo
 \fi
}%
\providecommand \natexlab [1]{#1}%
\providecommand \enquote  [1]{``#1''}%
\providecommand \bibnamefont  [1]{#1}%
\providecommand \bibfnamefont [1]{#1}%
\providecommand \citenamefont [1]{#1}%
\providecommand \href@noop [0]{\@secondoftwo}%
\providecommand \href [0]{\begingroup \@sanitize@url \@href}%
\providecommand \@href[1]{\@@startlink{#1}\@@href}%
\providecommand \@@href[1]{\endgroup#1\@@endlink}%
\providecommand \@sanitize@url [0]{\catcode `\\12\catcode `\$12\catcode
  `\&12\catcode `\#12\catcode `\^12\catcode `\_12\catcode `\%12\relax}%
\providecommand \@@startlink[1]{}%
\providecommand \@@endlink[0]{}%
\providecommand \url  [0]{\begingroup\@sanitize@url \@url }%
\providecommand \@url [1]{\endgroup\@href {#1}{\urlprefix }}%
\providecommand \urlprefix  [0]{URL }%
\providecommand \Eprint [0]{\href }%
\providecommand \doibase [0]{http://dx.doi.org/}%
\providecommand \selectlanguage [0]{\@gobble}%
\providecommand \bibinfo  [0]{\@secondoftwo}%
\providecommand \bibfield  [0]{\@secondoftwo}%
\providecommand \translation [1]{[#1]}%
\providecommand \BibitemOpen [0]{}%
\providecommand \bibitemStop [0]{}%
\providecommand \bibitemNoStop [0]{.\EOS\space}%
\providecommand \EOS [0]{\spacefactor3000\relax}%
\providecommand \BibitemShut  [1]{\csname bibitem#1\endcsname}%
\let\auto@bib@innerbib\@empty
\bibitem [{\citenamefont {Campisi}\ \emph {et~al.}(2011)\citenamefont
  {Campisi}, \citenamefont {H{\"a}nggi},\ and\ \citenamefont
  {Talkner}}]{campisi2011}%
  \BibitemOpen
  \bibfield  {author} {\bibinfo {author} {\bibfnamefont {Michele}\ \bibnamefont
  {Campisi}}, \bibinfo {author} {\bibfnamefont {Peter}\ \bibnamefont
  {H{\"a}nggi}}, \ and\ \bibinfo {author} {\bibfnamefont {Peter}\ \bibnamefont
  {Talkner}},\ }\bibfield  {title} {\enquote {\bibinfo {title} {Colloquium:
  Quantum fluctuation relations: Foundations and applications},}\ }\href
  {https://link.aps.org/doi/10.1103/RevModPhys.83.771} {\bibfield  {journal}
  {\bibinfo  {journal} {Rev. Mod. Phys.}\ }\textbf {\bibinfo {volume} {83}},\
  \bibinfo {pages} {771} (\bibinfo {year} {2011})}\BibitemShut {NoStop}%
\bibitem [{\citenamefont {Alicki}(1979)}]{Alicki1979}%
  \BibitemOpen
  \bibfield  {author} {\bibinfo {author} {\bibfnamefont {R}~\bibnamefont
  {Alicki}},\ }\bibfield  {title} {\enquote {\bibinfo {title} {The quantum open
  system as a model of the heat engine},}\ }\href {\doibase
  10.1088/0305-4470/12/5/007} {\bibfield  {journal} {\bibinfo  {journal} {J.
  Phys. A}\ }\textbf {\bibinfo {volume} {12}},\ \bibinfo {pages} {L103--L107}
  (\bibinfo {year} {1979})}\BibitemShut {NoStop}%
\bibitem [{\citenamefont {Kosloff}(2013)}]{Kosloff2013}%
  \BibitemOpen
  \bibfield  {author} {\bibinfo {author} {\bibfnamefont {Ronnie}\ \bibnamefont
  {Kosloff}},\ }\bibfield  {title} {\enquote {\bibinfo {title} {Quantum
  thermodynamics: A dynamical viewpoint},}\ }\href {\doibase 10.3390/e15062100}
  {\bibfield  {journal} {\bibinfo  {journal} {Entropy}\ }\textbf {\bibinfo
  {volume} {15}},\ \bibinfo {pages} {2100--2128} (\bibinfo {year}
  {2013})}\BibitemShut {NoStop}%
\bibitem [{\citenamefont {Frenzel}\ \emph {et~al.}(2014)\citenamefont
  {Frenzel}, \citenamefont {Jennings},\ and\ \citenamefont
  {Rudolph}}]{Frenzel2014}%
  \BibitemOpen
  \bibfield  {author} {\bibinfo {author} {\bibfnamefont {Max~F.}\ \bibnamefont
  {Frenzel}}, \bibinfo {author} {\bibfnamefont {David}\ \bibnamefont
  {Jennings}}, \ and\ \bibinfo {author} {\bibfnamefont {Terry}\ \bibnamefont
  {Rudolph}},\ }\bibfield  {title} {\enquote {\bibinfo {title} {Reexamination
  of pure qubit work extraction},}\ }\href {\doibase
  10.1103/PhysRevE.90.052136} {\bibfield  {journal} {\bibinfo  {journal} {Phys.
  Rev. E}\ }\textbf {\bibinfo {volume} {90}},\ \bibinfo {pages} {052136}
  (\bibinfo {year} {2014})}\BibitemShut {NoStop}%
\bibitem [{\citenamefont {Dahlsten}\ \emph {et~al.}(2017)\citenamefont
  {Dahlsten}, \citenamefont {Choi}, \citenamefont {Braun}, \citenamefont
  {Garner}, \citenamefont {Halpern},\ and\ \citenamefont
  {Vedral}}]{Dahlsten2017}%
  \BibitemOpen
  \bibfield  {author} {\bibinfo {author} {\bibfnamefont {Oscar~CO}\
  \bibnamefont {Dahlsten}}, \bibinfo {author} {\bibfnamefont {Mahn-Soo}\
  \bibnamefont {Choi}}, \bibinfo {author} {\bibfnamefont {Daniel}\ \bibnamefont
  {Braun}}, \bibinfo {author} {\bibfnamefont {Andrew~JP}\ \bibnamefont
  {Garner}}, \bibinfo {author} {\bibfnamefont {Nicole~Yunger}\ \bibnamefont
  {Halpern}}, \ and\ \bibinfo {author} {\bibfnamefont {Vlatko}\ \bibnamefont
  {Vedral}},\ }\bibfield  {title} {\enquote {\bibinfo {title} {Entropic
  equality for worst-case work at any protocol speed},}\ }\href
  {https://doi.org/10.1088/1367-2630/aa62ba} {\bibfield  {journal} {\bibinfo
  {journal} {New J. Phys.}\ }\textbf {\bibinfo {volume} {19}},\ \bibinfo
  {pages} {043013} (\bibinfo {year} {2017})}\BibitemShut {NoStop}%
\bibitem [{\citenamefont {Deffner}\ \emph {et~al.}(2016)\citenamefont
  {Deffner}, \citenamefont {Paz},\ and\ \citenamefont {Zurek}}]{Deffner2016}%
  \BibitemOpen
  \bibfield  {author} {\bibinfo {author} {\bibfnamefont {Sebastian}\
  \bibnamefont {Deffner}}, \bibinfo {author} {\bibfnamefont {Juan~Pablo}\
  \bibnamefont {Paz}}, \ and\ \bibinfo {author} {\bibfnamefont {Wojciech~H.}\
  \bibnamefont {Zurek}},\ }\bibfield  {title} {\enquote {\bibinfo {title}
  {Quantum work and the thermodynamic cost of quantum measurements},}\ }\href
  {\doibase 10.1103/physreve.94.010103} {\bibfield  {journal} {\bibinfo
  {journal} {Phys. Rev. E}\ }\textbf {\bibinfo {volume} {94}},\ \bibinfo
  {pages} {010103} (\bibinfo {year} {2016})}\BibitemShut {NoStop}%
\bibitem [{\citenamefont {Alipour}\ \emph {et~al.}(2022)\citenamefont
  {Alipour}, \citenamefont {Rezakhani}, \citenamefont {Chenu}, \citenamefont
  {Del~Campo},\ and\ \citenamefont {Ala-Nissila}}]{Alipour2022}%
  \BibitemOpen
  \bibfield  {author} {\bibinfo {author} {\bibfnamefont {S}~\bibnamefont
  {Alipour}}, \bibinfo {author} {\bibfnamefont {AT}~\bibnamefont {Rezakhani}},
  \bibinfo {author} {\bibfnamefont {A}~\bibnamefont {Chenu}}, \bibinfo {author}
  {\bibfnamefont {A}~\bibnamefont {Del~Campo}}, \ and\ \bibinfo {author}
  {\bibfnamefont {Tapio}\ \bibnamefont {Ala-Nissila}},\ }\bibfield  {title}
  {\enquote {\bibinfo {title} {Entropy-based formulation of thermodynamics in
  arbitrary quantum evolution},}\ }\href
  {https://link.aps.org/doi/10.1103/PhysRevA.105.L040201} {\bibfield  {journal}
  {\bibinfo  {journal} {Phys. Rev. A}\ }\textbf {\bibinfo {volume} {105}},\
  \bibinfo {pages} {L040201} (\bibinfo {year} {2022})}\BibitemShut {NoStop}%
\bibitem [{\citenamefont {Talkner}\ \emph {et~al.}(2007)\citenamefont
  {Talkner}, \citenamefont {Lutz},\ and\ \citenamefont
  {H\"anggi}}]{Talkner2007}%
  \BibitemOpen
  \bibfield  {author} {\bibinfo {author} {\bibfnamefont {Peter}\ \bibnamefont
  {Talkner}}, \bibinfo {author} {\bibfnamefont {Eric}\ \bibnamefont {Lutz}}, \
  and\ \bibinfo {author} {\bibfnamefont {Peter}\ \bibnamefont {H\"anggi}},\
  }\bibfield  {title} {\enquote {\bibinfo {title} {Fluctuation theorems: Work
  is not an observable},}\ }\href {\doibase 10.1103/PhysRevE.75.050102}
  {\bibfield  {journal} {\bibinfo  {journal} {Phys. Rev. E}\ }\textbf {\bibinfo
  {volume} {75}},\ \bibinfo {pages} {050102} (\bibinfo {year}
  {2007})}\BibitemShut {NoStop}%
\bibitem [{\citenamefont {Deffner}\ and\ \citenamefont
  {Lutz}(2008)}]{Deffner2008}%
  \BibitemOpen
  \bibfield  {author} {\bibinfo {author} {\bibfnamefont {S.}~\bibnamefont
  {Deffner}}\ and\ \bibinfo {author} {\bibfnamefont {E.}~\bibnamefont {Lutz}},\
  }\bibfield  {title} {\enquote {\bibinfo {title} {Nonequilibrium work
  distribution of a quantum harmonic oscillator},}\ }\href {\doibase
  10.1103/physreve.77.021128} {\bibfield  {journal} {\bibinfo  {journal} {Phys.
  Rev. E}\ }\textbf {\bibinfo {volume} {77}},\ \bibinfo {pages} {021128}
  (\bibinfo {year} {2008})}\BibitemShut {NoStop}%
\bibitem [{\citenamefont {Goold}\ \emph {et~al.}(2016)\citenamefont {Goold},
  \citenamefont {Huber}, \citenamefont {Riera}, \citenamefont {Del~Rio},\ and\
  \citenamefont {Skrzypczyk}}]{Goold2016}%
  \BibitemOpen
  \bibfield  {author} {\bibinfo {author} {\bibfnamefont {John}\ \bibnamefont
  {Goold}}, \bibinfo {author} {\bibfnamefont {Marcus}\ \bibnamefont {Huber}},
  \bibinfo {author} {\bibfnamefont {Arnau}\ \bibnamefont {Riera}}, \bibinfo
  {author} {\bibfnamefont {L{\'\i}dia}\ \bibnamefont {Del~Rio}}, \ and\
  \bibinfo {author} {\bibfnamefont {Paul}\ \bibnamefont {Skrzypczyk}},\
  }\bibfield  {title} {\enquote {\bibinfo {title} {The role of quantum
  information in thermodynamics—a topical review},}\ }\href
  {https://doi.org/10.1088/1751-8113/49/14/143001} {\bibfield  {journal}
  {\bibinfo  {journal} {J. Phys. A}\ }\textbf {\bibinfo {volume} {49}},\
  \bibinfo {pages} {143001} (\bibinfo {year} {2016})}\BibitemShut {NoStop}%
\bibitem [{\citenamefont {Batalh\~ao}\ \emph {et~al.}(2014)\citenamefont
  {Batalh\~ao}, \citenamefont {Souza}, \citenamefont {Mazzola}, \citenamefont
  {Auccaise}, \citenamefont {Sarthour}, \citenamefont {Oliveira}, \citenamefont
  {Goold}, \citenamefont {De~Chiara}, \citenamefont {Paternostro},\ and\
  \citenamefont {Serra}}]{Batalhao2014}%
  \BibitemOpen
  \bibfield  {author} {\bibinfo {author} {\bibfnamefont {Tiago~B.}\
  \bibnamefont {Batalh\~ao}}, \bibinfo {author} {\bibfnamefont {Alexandre~M.}\
  \bibnamefont {Souza}}, \bibinfo {author} {\bibfnamefont {Laura}\ \bibnamefont
  {Mazzola}}, \bibinfo {author} {\bibfnamefont {Ruben}\ \bibnamefont
  {Auccaise}}, \bibinfo {author} {\bibfnamefont {Roberto~S.}\ \bibnamefont
  {Sarthour}}, \bibinfo {author} {\bibfnamefont {Ivan~S.}\ \bibnamefont
  {Oliveira}}, \bibinfo {author} {\bibfnamefont {John}\ \bibnamefont {Goold}},
  \bibinfo {author} {\bibfnamefont {Gabriele}\ \bibnamefont {De~Chiara}},
  \bibinfo {author} {\bibfnamefont {Mauro}\ \bibnamefont {Paternostro}}, \ and\
  \bibinfo {author} {\bibfnamefont {Roberto~M.}\ \bibnamefont {Serra}},\
  }\bibfield  {title} {\enquote {\bibinfo {title} {Experimental reconstruction
  of work distribution and study of fluctuation relations in a closed quantum
  system},}\ }\href {\doibase 10.1103/PhysRevLett.113.140601} {\bibfield
  {journal} {\bibinfo  {journal} {Phys. Rev. Lett.}\ }\textbf {\bibinfo
  {volume} {113}},\ \bibinfo {pages} {140601} (\bibinfo {year}
  {2014})}\BibitemShut {NoStop}%
\bibitem [{\citenamefont {De~Chiara}\ \emph {et~al.}(2015)\citenamefont
  {De~Chiara}, \citenamefont {Roncaglia},\ and\ \citenamefont
  {Paz}}]{DeChiara2015}%
  \BibitemOpen
  \bibfield  {author} {\bibinfo {author} {\bibfnamefont {Gabriele}\
  \bibnamefont {De~Chiara}}, \bibinfo {author} {\bibfnamefont {Augusto~J}\
  \bibnamefont {Roncaglia}}, \ and\ \bibinfo {author} {\bibfnamefont
  {Juan~Pablo}\ \bibnamefont {Paz}},\ }\bibfield  {title} {\enquote {\bibinfo
  {title} {Measuring work and heat in ultracold quantum gases},}\ }\href
  {https://doi.org/10.1088/1367-2630/17/3/035004} {\bibfield  {journal}
  {\bibinfo  {journal} {New J. Phys.}\ }\textbf {\bibinfo {volume} {17}},\
  \bibinfo {pages} {035004} (\bibinfo {year} {2015})}\BibitemShut {NoStop}%
\bibitem [{\citenamefont {Campisi}\ and\ \citenamefont
  {Goold}(2017)}]{Campisi2017}%
  \BibitemOpen
  \bibfield  {author} {\bibinfo {author} {\bibfnamefont {Michele}\ \bibnamefont
  {Campisi}}\ and\ \bibinfo {author} {\bibfnamefont {John}\ \bibnamefont
  {Goold}},\ }\bibfield  {title} {\enquote {\bibinfo {title} {Thermodynamics of
  quantum information scrambling},}\ }\href {\doibase
  10.1103/PhysRevE.95.062127} {\bibfield  {journal} {\bibinfo  {journal} {Phys.
  Rev. E}\ }\textbf {\bibinfo {volume} {95}},\ \bibinfo {pages} {062127}
  (\bibinfo {year} {2017})}\BibitemShut {NoStop}%
\bibitem [{\citenamefont {Chenu}\ \emph {et~al.}(2018)\citenamefont {Chenu},
  \citenamefont {Egusquiza}, \citenamefont {Molina-Vilaplana},\ and\
  \citenamefont {del Campo}}]{ChenuSciRep}%
  \BibitemOpen
  \bibfield  {author} {\bibinfo {author} {\bibfnamefont {A.}~\bibnamefont
  {Chenu}}, \bibinfo {author} {\bibfnamefont {I.~L.}\ \bibnamefont
  {Egusquiza}}, \bibinfo {author} {\bibfnamefont {J.}~\bibnamefont
  {Molina-Vilaplana}}, \ and\ \bibinfo {author} {\bibfnamefont
  {A.}~\bibnamefont {del Campo}},\ }\bibfield  {title} {\enquote {\bibinfo
  {title} {Quantum work statistics, loschmidt echo and information
  scrambling},}\ }\href {\doibase 10.1038/s41598-018-30982-w} {\bibfield
  {journal} {\bibinfo  {journal} {Sci. Rep.}\ }\textbf {\bibinfo {volume}
  {8}},\ \bibinfo {pages} {12634} (\bibinfo {year} {2018})}\BibitemShut
  {NoStop}%
\bibitem [{\citenamefont {Chenu}\ \emph {et~al.}(2019)\citenamefont {Chenu},
  \citenamefont {Molina-Vilaplana},\ and\ \citenamefont {del
  Campo}}]{ChenuQuantum}%
  \BibitemOpen
  \bibfield  {author} {\bibinfo {author} {\bibfnamefont {A.}~\bibnamefont
  {Chenu}}, \bibinfo {author} {\bibfnamefont {J.}~\bibnamefont
  {Molina-Vilaplana}}, \ and\ \bibinfo {author} {\bibfnamefont
  {A.}~\bibnamefont {del Campo}},\ }\bibfield  {title} {\enquote {\bibinfo
  {title} {Work statistics, loschmidt echo and information scrambling in
  chaotic quantum systems},}\ }\href {\doibase 10.22331/q-2019-03-04-127}
  {\bibfield  {journal} {\bibinfo  {journal} {Quantum}\ }\textbf {\bibinfo
  {volume} {3}},\ \bibinfo {pages} {127} (\bibinfo {year} {2019})}\BibitemShut
  {NoStop}%
\bibitem [{\citenamefont {del Campo}(2018)}]{delCampoPRL}%
  \BibitemOpen
  \bibfield  {author} {\bibinfo {author} {\bibfnamefont {A.}~\bibnamefont {del
  Campo}},\ }\bibfield  {title} {\enquote {\bibinfo {title} {Universal
  statistics of topological defects formed in a quantum phase transition},}\
  }\href {https://doi.org/10.1103/physrevlett.121.200601} {\bibfield  {journal}
  {\bibinfo  {journal} {Physical Review Letters}\ }\textbf {\bibinfo {volume}
  {121}} (\bibinfo {year} {2018})}\BibitemShut {NoStop}%
\bibitem [{\citenamefont {Fei}\ \emph {et~al.}(2020)\citenamefont {Fei},
  \citenamefont {Freitas}, \citenamefont {Cavina}, \citenamefont {Quan},\ and\
  \citenamefont {Esposito}}]{Fei2020}%
  \BibitemOpen
  \bibfield  {author} {\bibinfo {author} {\bibfnamefont {Zhaoyu}\ \bibnamefont
  {Fei}}, \bibinfo {author} {\bibfnamefont {Nahuel}\ \bibnamefont {Freitas}},
  \bibinfo {author} {\bibfnamefont {Vasco}\ \bibnamefont {Cavina}}, \bibinfo
  {author} {\bibfnamefont {H.~T.}\ \bibnamefont {Quan}}, \ and\ \bibinfo
  {author} {\bibfnamefont {Massimiliano}\ \bibnamefont {Esposito}},\ }\bibfield
   {title} {\enquote {\bibinfo {title} {Work statistics across a quantum phase
  transition},}\ }\href {\doibase 10.1103/PhysRevLett.124.170603} {\bibfield
  {journal} {\bibinfo  {journal} {Phys. Rev. Lett.}\ }\textbf {\bibinfo
  {volume} {124}},\ \bibinfo {pages} {170603} (\bibinfo {year}
  {2020})}\BibitemShut {NoStop}%
\bibitem [{\citenamefont {Fusco}\ \emph {et~al.}(2014)\citenamefont {Fusco},
  \citenamefont {Pigeon}, \citenamefont {Apollaro}, \citenamefont {Xuereb},
  \citenamefont {Mazzola}, \citenamefont {Campisi}, \citenamefont {Ferraro},
  \citenamefont {Paternostro},\ and\ \citenamefont {De~Chiara}}]{Fusco2014}%
  \BibitemOpen
  \bibfield  {author} {\bibinfo {author} {\bibfnamefont {L.}~\bibnamefont
  {Fusco}}, \bibinfo {author} {\bibfnamefont {S.}~\bibnamefont {Pigeon}},
  \bibinfo {author} {\bibfnamefont {T.~J.~G.}\ \bibnamefont {Apollaro}},
  \bibinfo {author} {\bibfnamefont {A.}~\bibnamefont {Xuereb}}, \bibinfo
  {author} {\bibfnamefont {L.}~\bibnamefont {Mazzola}}, \bibinfo {author}
  {\bibfnamefont {M.}~\bibnamefont {Campisi}}, \bibinfo {author} {\bibfnamefont
  {A.}~\bibnamefont {Ferraro}}, \bibinfo {author} {\bibfnamefont
  {M.}~\bibnamefont {Paternostro}}, \ and\ \bibinfo {author} {\bibfnamefont
  {G.}~\bibnamefont {De~Chiara}},\ }\bibfield  {title} {\enquote {\bibinfo
  {title} {Assessing the nonequilibrium thermodynamics in a quenched quantum
  many-body system via single projective measurements},}\ }\href {\doibase
  10.1103/PhysRevX.4.031029} {\bibfield  {journal} {\bibinfo  {journal} {Phys.
  Rev. X}\ }\textbf {\bibinfo {volume} {4}},\ \bibinfo {pages} {031029}
  (\bibinfo {year} {2014})}\BibitemShut {NoStop}%
\bibitem [{\citenamefont {Zawadzki}\ \emph {et~al.}(2023)\citenamefont
  {Zawadzki}, \citenamefont {Kiely}, \citenamefont {Landi},\ and\ \citenamefont
  {Campbell}}]{Zawadzki2022}%
  \BibitemOpen
  \bibfield  {author} {\bibinfo {author} {\bibfnamefont {Krissia}\ \bibnamefont
  {Zawadzki}}, \bibinfo {author} {\bibfnamefont {Anthony}\ \bibnamefont
  {Kiely}}, \bibinfo {author} {\bibfnamefont {Gabriel~T}\ \bibnamefont
  {Landi}}, \ and\ \bibinfo {author} {\bibfnamefont {Steve}\ \bibnamefont
  {Campbell}},\ }\bibfield  {title} {\enquote {\bibinfo {title} {Non-gaussian
  work statistics at finite-time driving},}\ }\href {\doibase
  10.1103/PhysRevA.107.012209} {\bibfield  {journal} {\bibinfo  {journal}
  {Phys. Rev. A}\ }\textbf {\bibinfo {volume} {107}},\ \bibinfo {pages}
  {012209} (\bibinfo {year} {2023})}\BibitemShut {NoStop}%
\bibitem [{\citenamefont {Zawadzki}\ \emph {et~al.}(2020)\citenamefont
  {Zawadzki}, \citenamefont {Serra},\ and\ \citenamefont
  {D'Amico}}]{Zawadzki2020}%
  \BibitemOpen
  \bibfield  {author} {\bibinfo {author} {\bibfnamefont {Krissia}\ \bibnamefont
  {Zawadzki}}, \bibinfo {author} {\bibfnamefont {Roberto~M.}\ \bibnamefont
  {Serra}}, \ and\ \bibinfo {author} {\bibfnamefont {Irene}\ \bibnamefont
  {D'Amico}},\ }\bibfield  {title} {\enquote {\bibinfo {title}
  {Work-distribution quantumness and irreversibility when crossing a quantum
  phase transition in finite time},}\ }\href {\doibase
  10.1103/PhysRevResearch.2.033167} {\bibfield  {journal} {\bibinfo  {journal}
  {Phys. Rev. Research}\ }\textbf {\bibinfo {volume} {2}},\ \bibinfo {pages}
  {033167} (\bibinfo {year} {2020})}\BibitemShut {NoStop}%
\bibitem [{\citenamefont {Juan-Delgado}\ and\ \citenamefont
  {Chenu}(2021)}]{JuanDelgado2021}%
  \BibitemOpen
  \bibfield  {author} {\bibinfo {author} {\bibfnamefont {Adri\'an}\
  \bibnamefont {Juan-Delgado}}\ and\ \bibinfo {author} {\bibfnamefont
  {Aur\'elia}\ \bibnamefont {Chenu}},\ }\bibfield  {title} {\enquote {\bibinfo
  {title} {First law of quantum thermodynamics in a driven open two-level
  system},}\ }\href {\doibase 10.1103/PhysRevA.104.022219} {\bibfield
  {journal} {\bibinfo  {journal} {Phys. Rev. A}\ }\textbf {\bibinfo {volume}
  {104}},\ \bibinfo {pages} {022219} (\bibinfo {year} {2021})}\BibitemShut
  {NoStop}%
\bibitem [{\citenamefont {D\'{i}az}\ \emph {et~al.}(2020)\citenamefont
  {D\'{i}az}, \citenamefont {Guarnieri},\ and\ \citenamefont
  {Paternostro}}]{Diaz2020}%
  \BibitemOpen
  \bibfield  {author} {\bibinfo {author} {\bibfnamefont {Mar\'{i}a~Garc\'{i}a}\
  \bibnamefont {D\'{i}az}}, \bibinfo {author} {\bibfnamefont {Giacomo}\
  \bibnamefont {Guarnieri}}, \ and\ \bibinfo {author} {\bibfnamefont {Mauro}\
  \bibnamefont {Paternostro}},\ }\bibfield  {title} {\enquote {\bibinfo {title}
  {Quantum work statistics with initial coherence},}\ }\href
  {https://www.mdpi.com/1099-4300/22/11/1223} {\bibfield  {journal} {\bibinfo
  {journal} {Entropy}\ }\textbf {\bibinfo {volume} {22}} (\bibinfo {year}
  {2020})}\BibitemShut {NoStop}%
\bibitem [{\citenamefont {Korzekwa}\ \emph {et~al.}(2016)\citenamefont
  {Korzekwa}, \citenamefont {Lostaglio}, \citenamefont {Oppenheim},\ and\
  \citenamefont {Jennings}}]{Korzekwa2016}%
  \BibitemOpen
  \bibfield  {author} {\bibinfo {author} {\bibfnamefont {Kamil}\ \bibnamefont
  {Korzekwa}}, \bibinfo {author} {\bibfnamefont {Matteo}\ \bibnamefont
  {Lostaglio}}, \bibinfo {author} {\bibfnamefont {Jonathan}\ \bibnamefont
  {Oppenheim}}, \ and\ \bibinfo {author} {\bibfnamefont {David}\ \bibnamefont
  {Jennings}},\ }\bibfield  {title} {\enquote {\bibinfo {title} {The extraction
  of work from quantum coherence},}\ }\href {\doibase
  10.1088/1367-2630/18/2/023045} {\bibfield  {journal} {\bibinfo  {journal}
  {New J. Phys.}\ }\textbf {\bibinfo {volume} {18}},\ \bibinfo {pages} {023045}
  (\bibinfo {year} {2016})}\BibitemShut {NoStop}%
\bibitem [{\citenamefont {Goold}\ \emph {et~al.}(2018)\citenamefont {Goold},
  \citenamefont {Plastina}, \citenamefont {Gambassi},\ and\ \citenamefont
  {Silva}}]{goold2018}%
  \BibitemOpen
  \bibfield  {author} {\bibinfo {author} {\bibfnamefont {John}\ \bibnamefont
  {Goold}}, \bibinfo {author} {\bibfnamefont {Francesco}\ \bibnamefont
  {Plastina}}, \bibinfo {author} {\bibfnamefont {Andrea}\ \bibnamefont
  {Gambassi}}, \ and\ \bibinfo {author} {\bibfnamefont {Alessandro}\
  \bibnamefont {Silva}},\ }\bibfield  {title} {\enquote {\bibinfo {title} {The
  role of quantum work statistics in many-body physics},}\ }\href
  {https://link.springer.com/chapter/10.1007/978-3-319-99046-0_13} {\bibfield
  {journal} {\bibinfo  {journal} {Thermodynamics in the Quantum Regime}\ ,\
  \bibinfo {pages} {317--336}} (\bibinfo {year} {2018})}\BibitemShut {NoStop}%
\bibitem [{\citenamefont {Klatzow}\ \emph {et~al.}(2019)\citenamefont
  {Klatzow}, \citenamefont {Becker}, \citenamefont {Ledingham}, \citenamefont
  {Weinzetl}, \citenamefont {Kaczmarek}, \citenamefont {Saunders},
  \citenamefont {Nunn}, \citenamefont {Walmsley}, \citenamefont {Uzdin},\ and\
  \citenamefont {Poem}}]{Klatzow2019}%
  \BibitemOpen
  \bibfield  {author} {\bibinfo {author} {\bibfnamefont {James}\ \bibnamefont
  {Klatzow}}, \bibinfo {author} {\bibfnamefont {Jonas~N.}\ \bibnamefont
  {Becker}}, \bibinfo {author} {\bibfnamefont {Patrick~M.}\ \bibnamefont
  {Ledingham}}, \bibinfo {author} {\bibfnamefont {Christian}\ \bibnamefont
  {Weinzetl}}, \bibinfo {author} {\bibfnamefont {Krzysztof~T.}\ \bibnamefont
  {Kaczmarek}}, \bibinfo {author} {\bibfnamefont {Dylan~J.}\ \bibnamefont
  {Saunders}}, \bibinfo {author} {\bibfnamefont {Joshua}\ \bibnamefont {Nunn}},
  \bibinfo {author} {\bibfnamefont {Ian~A.}\ \bibnamefont {Walmsley}}, \bibinfo
  {author} {\bibfnamefont {Raam}\ \bibnamefont {Uzdin}}, \ and\ \bibinfo
  {author} {\bibfnamefont {Eilon}\ \bibnamefont {Poem}},\ }\bibfield  {title}
  {\enquote {\bibinfo {title} {Experimental demonstration of quantum effects in
  the operation of microscopic heat engines},}\ }\href {\doibase
  10.1103/PhysRevLett.122.110601} {\bibfield  {journal} {\bibinfo  {journal}
  {Phys. Rev. Lett.}\ }\textbf {\bibinfo {volume} {122}},\ \bibinfo {pages}
  {110601} (\bibinfo {year} {2019})}\BibitemShut {NoStop}%
\bibitem [{\citenamefont {Huber}\ \emph {et~al.}(2015)\citenamefont {Huber},
  \citenamefont {Perarnau-Llobet}, \citenamefont {Hovhannisyan}, \citenamefont
  {Skrzypczyk}, \citenamefont {Kl{\"o}ckl}, \citenamefont {Brunner},\ and\
  \citenamefont {Ac{\'\i}n}}]{Huber2015}%
  \BibitemOpen
  \bibfield  {author} {\bibinfo {author} {\bibfnamefont {Marcus}\ \bibnamefont
  {Huber}}, \bibinfo {author} {\bibfnamefont {Mart{\'\i}}\ \bibnamefont
  {Perarnau-Llobet}}, \bibinfo {author} {\bibfnamefont {Karen~V}\ \bibnamefont
  {Hovhannisyan}}, \bibinfo {author} {\bibfnamefont {Paul}\ \bibnamefont
  {Skrzypczyk}}, \bibinfo {author} {\bibfnamefont {Claude}\ \bibnamefont
  {Kl{\"o}ckl}}, \bibinfo {author} {\bibfnamefont {Nicolas}\ \bibnamefont
  {Brunner}}, \ and\ \bibinfo {author} {\bibfnamefont {Antonio}\ \bibnamefont
  {Ac{\'\i}n}},\ }\bibfield  {title} {\enquote {\bibinfo {title} {Thermodynamic
  cost of creating correlations},}\ }\href
  {https://doi.org/10.1088/1367-2630/17/6/065008} {\bibfield  {journal}
  {\bibinfo  {journal} {New J. Phys.}\ }\textbf {\bibinfo {volume} {17}},\
  \bibinfo {pages} {065008} (\bibinfo {year} {2015})}\BibitemShut {NoStop}%
\bibitem [{\citenamefont {Misra}\ \emph {et~al.}(2016)\citenamefont {Misra},
  \citenamefont {Singh}, \citenamefont {Bhattacharya},\ and\ \citenamefont
  {Pati}}]{Misra2016}%
  \BibitemOpen
  \bibfield  {author} {\bibinfo {author} {\bibfnamefont {Avijit}\ \bibnamefont
  {Misra}}, \bibinfo {author} {\bibfnamefont {Uttam}\ \bibnamefont {Singh}},
  \bibinfo {author} {\bibfnamefont {Samyadeb}\ \bibnamefont {Bhattacharya}}, \
  and\ \bibinfo {author} {\bibfnamefont {Arun~Kumar}\ \bibnamefont {Pati}},\
  }\bibfield  {title} {\enquote {\bibinfo {title} {Energy cost of creating
  quantum coherence},}\ }\href
  {https://link.aps.org/doi/10.1103/PhysRevA.93.052335} {\bibfield  {journal}
  {\bibinfo  {journal} {Phys. Rev. A}\ }\textbf {\bibinfo {volume} {93}},\
  \bibinfo {pages} {052335} (\bibinfo {year} {2016})}\BibitemShut {NoStop}%
\bibitem [{\citenamefont {Miller}\ \emph {et~al.}(2020)\citenamefont {Miller},
  \citenamefont {Guarnieri}, \citenamefont {Mitchison},\ and\ \citenamefont
  {Goold}}]{Miller2020}%
  \BibitemOpen
  \bibfield  {author} {\bibinfo {author} {\bibfnamefont {Harry J.~D.}\
  \bibnamefont {Miller}}, \bibinfo {author} {\bibfnamefont {Giacomo}\
  \bibnamefont {Guarnieri}}, \bibinfo {author} {\bibfnamefont {Mark~T.}\
  \bibnamefont {Mitchison}}, \ and\ \bibinfo {author} {\bibfnamefont {John}\
  \bibnamefont {Goold}},\ }\bibfield  {title} {\enquote {\bibinfo {title}
  {Quantum fluctuations hinder finite-time information erasure near the
  landauer limit},}\ }\href
  {https://link.aps.org/doi/10.1103/PhysRevLett.125.160602} {\bibfield
  {journal} {\bibinfo  {journal} {Phys. Rev. Lett.}\ }\textbf {\bibinfo
  {volume} {125}},\ \bibinfo {pages} {160602} (\bibinfo {year}
  {2020})}\BibitemShut {NoStop}%
\bibitem [{\citenamefont {Santos}\ \emph {et~al.}(2019)\citenamefont {Santos},
  \citenamefont {C{\'e}leri}, \citenamefont {Landi},\ and\ \citenamefont
  {Paternostro}}]{Santos2019}%
  \BibitemOpen
  \bibfield  {author} {\bibinfo {author} {\bibfnamefont {Jader~P}\ \bibnamefont
  {Santos}}, \bibinfo {author} {\bibfnamefont {Lucas~C}\ \bibnamefont
  {C{\'e}leri}}, \bibinfo {author} {\bibfnamefont {Gabriel~T}\ \bibnamefont
  {Landi}}, \ and\ \bibinfo {author} {\bibfnamefont {Mauro}\ \bibnamefont
  {Paternostro}},\ }\bibfield  {title} {\enquote {\bibinfo {title} {The role of
  quantum coherence in non-equilibrium entropy production},}\ }\href@noop {}
  {\bibfield  {journal} {\bibinfo  {journal} {npj Quantum Information}\
  }\textbf {\bibinfo {volume} {5}},\ \bibinfo {pages} {1--7} (\bibinfo {year}
  {2019})}\BibitemShut {NoStop}%
\bibitem [{\citenamefont {Francica}\ \emph {et~al.}(2019)\citenamefont
  {Francica}, \citenamefont {Goold},\ and\ \citenamefont
  {Plastina}}]{Francica2019}%
  \BibitemOpen
  \bibfield  {author} {\bibinfo {author} {\bibfnamefont {G.}~\bibnamefont
  {Francica}}, \bibinfo {author} {\bibfnamefont {J.}~\bibnamefont {Goold}}, \
  and\ \bibinfo {author} {\bibfnamefont {F.}~\bibnamefont {Plastina}},\
  }\bibfield  {title} {\enquote {\bibinfo {title} {Role of coherence in the
  nonequilibrium thermodynamics of quantum systems},}\ }\href
  {https://doi.org/10.1103/physreve.99.042105} {\bibfield  {journal} {\bibinfo
  {journal} {Physical Review E}\ }\textbf {\bibinfo {volume} {99}} (\bibinfo
  {year} {2019})}\BibitemShut {NoStop}%
\bibitem [{\citenamefont {Salazar}(2021)}]{Salazar2021}%
  \BibitemOpen
  \bibfield  {author} {\bibinfo {author} {\bibfnamefont {Domingos S.~P.}\
  \bibnamefont {Salazar}},\ }\bibfield  {title} {\enquote {\bibinfo {title}
  {Information bound for entropy production from the detailed fluctuation
  theorem},}\ }\href {https://link.aps.org/doi/10.1103/PhysRevE.103.022122}
  {\bibfield  {journal} {\bibinfo  {journal} {Phys. Rev. E}\ }\textbf {\bibinfo
  {volume} {103}},\ \bibinfo {pages} {022122} (\bibinfo {year}
  {2021})}\BibitemShut {NoStop}%
\bibitem [{Note1()}]{Note1}%
  \BibitemOpen
  \bibinfo {note} {The work distribution~\protect \textup {\hbox {\mathsurround
  \z@ \protect \normalfont (\ignorespaces \ref {PW}\unskip \@@italiccorr )}} is
  usually defined with a Dirac delta. As far as the entropy is concerned,
  however, the discreteness of the support is important when dealing with the
  entropy, which is why we defined it here with a Kronecker delta
  instead.}\BibitemShut {Stop}%
\bibitem [{Sup()}]{SupMat}%
  \BibitemOpen
  \href@noop {} {\emph {\bibinfo {title} {{See Supplemental
  Material}}}}\BibitemShut {NoStop}%
\bibitem [{\citenamefont {Baumgratz}\ \emph {et~al.}(2014)\citenamefont
  {Baumgratz}, \citenamefont {Cramer},\ and\ \citenamefont
  {Plenio}}]{Baumgratz2014}%
  \BibitemOpen
  \bibfield  {author} {\bibinfo {author} {\bibfnamefont {T.}~\bibnamefont
  {Baumgratz}}, \bibinfo {author} {\bibfnamefont {M.}~\bibnamefont {Cramer}}, \
  and\ \bibinfo {author} {\bibfnamefont {M.~B.}\ \bibnamefont {Plenio}},\
  }\bibfield  {title} {\enquote {\bibinfo {title} {Quantifying coherence},}\
  }\href {\doibase 10.1103/PhysRevLett.113.140401} {\bibfield  {journal}
  {\bibinfo  {journal} {Phys. Rev. Lett.}\ }\textbf {\bibinfo {volume} {113}},\
  \bibinfo {pages} {140401} (\bibinfo {year} {2014})}\BibitemShut {NoStop}%
\bibitem [{\citenamefont {Pietracaprina}\ \emph {et~al.}(2017)\citenamefont
  {Pietracaprina}, \citenamefont {Gogolin},\ and\ \citenamefont
  {Goold}}]{Pietracaprina2017}%
  \BibitemOpen
  \bibfield  {author} {\bibinfo {author} {\bibfnamefont {F.}~\bibnamefont
  {Pietracaprina}}, \bibinfo {author} {\bibfnamefont {C.}~\bibnamefont
  {Gogolin}}, \ and\ \bibinfo {author} {\bibfnamefont {J.}~\bibnamefont
  {Goold}},\ }\bibfield  {title} {\enquote {\bibinfo {title} {Total
  correlations of the diagonal ensemble as a generic indicator for ergodicity
  breaking in quantum systems},}\ }\href {\doibase 10.1103/PhysRevB.95.125118}
  {\bibfield  {journal} {\bibinfo  {journal} {Phys. Rev. B}\ }\textbf {\bibinfo
  {volume} {95}},\ \bibinfo {pages} {125118} (\bibinfo {year}
  {2017})}\BibitemShut {NoStop}%
\bibitem [{\citenamefont {{\c{C}}akan}\ \emph {et~al.}(2021)\citenamefont
  {{\c{C}}akan}, \citenamefont {Cirac},\ and\ \citenamefont
  {Ba{\~n}uls}}]{Cakan2021}%
  \BibitemOpen
  \bibfield  {author} {\bibinfo {author} {\bibfnamefont {Asl{\i}}\ \bibnamefont
  {{\c{C}}akan}}, \bibinfo {author} {\bibfnamefont {J~Ignacio}\ \bibnamefont
  {Cirac}}, \ and\ \bibinfo {author} {\bibfnamefont {Mari~Carmen}\ \bibnamefont
  {Ba{\~n}uls}},\ }\bibfield  {title} {\enquote {\bibinfo {title}
  {Approximating the long time average of the density operator: Diagonal
  ensemble},}\ }\href {\doibase 10.1103/PhysRevB.103.115113} {\bibfield
  {journal} {\bibinfo  {journal} {Phys. Rev. B}\ }\textbf {\bibinfo {volume}
  {103}},\ \bibinfo {pages} {115113} (\bibinfo {year} {2021})}\BibitemShut
  {NoStop}%
\bibitem [{\citenamefont {Goold}\ \emph {et~al.}(2015)\citenamefont {Goold},
  \citenamefont {Gogolin}, \citenamefont {Clark}, \citenamefont {Eisert},
  \citenamefont {Scardicchio},\ and\ \citenamefont {Silva}}]{Goold2015}%
  \BibitemOpen
  \bibfield  {author} {\bibinfo {author} {\bibfnamefont {J.}~\bibnamefont
  {Goold}}, \bibinfo {author} {\bibfnamefont {C.}~\bibnamefont {Gogolin}},
  \bibinfo {author} {\bibfnamefont {S.~R.}\ \bibnamefont {Clark}}, \bibinfo
  {author} {\bibfnamefont {J.}~\bibnamefont {Eisert}}, \bibinfo {author}
  {\bibfnamefont {A.}~\bibnamefont {Scardicchio}}, \ and\ \bibinfo {author}
  {\bibfnamefont {A.}~\bibnamefont {Silva}},\ }\bibfield  {title} {\enquote
  {\bibinfo {title} {Total correlations of the diagonal ensemble herald the
  many-body localization transition},}\ }\href {\doibase
  10.1103/PhysRevB.92.180202} {\bibfield  {journal} {\bibinfo  {journal} {Phys.
  Rev. B}\ }\textbf {\bibinfo {volume} {92}},\ \bibinfo {pages} {180202}
  (\bibinfo {year} {2015})}\BibitemShut {NoStop}%
\bibitem [{\citenamefont {Mzaouali}\ \emph {et~al.}(2021)\citenamefont
  {Mzaouali}, \citenamefont {Puebla}, \citenamefont {Goold}, \citenamefont
  {El~Baz},\ and\ \citenamefont {Campbell}}]{Mzaouali2021}%
  \BibitemOpen
  \bibfield  {author} {\bibinfo {author} {\bibfnamefont {Zakaria}\ \bibnamefont
  {Mzaouali}}, \bibinfo {author} {\bibfnamefont {Ricardo}\ \bibnamefont
  {Puebla}}, \bibinfo {author} {\bibfnamefont {John}\ \bibnamefont {Goold}},
  \bibinfo {author} {\bibfnamefont {Morad}\ \bibnamefont {El~Baz}}, \ and\
  \bibinfo {author} {\bibfnamefont {Steve}\ \bibnamefont {Campbell}},\
  }\bibfield  {title} {\enquote {\bibinfo {title} {Work statistics and symmetry
  breaking in an excited-state quantum phase transition},}\ }\href {\doibase
  10.1103/PhysRevE.103.032145} {\bibfield  {journal} {\bibinfo  {journal}
  {Phys. Rev. E}\ }\textbf {\bibinfo {volume} {103}},\ \bibinfo {pages}
  {032145} (\bibinfo {year} {2021})}\BibitemShut {NoStop}%
\bibitem [{\citenamefont {Wang}\ and\ \citenamefont
  {P\'erez-Bernal}(2021)}]{Wang2021}%
  \BibitemOpen
  \bibfield  {author} {\bibinfo {author} {\bibfnamefont {Qian}\ \bibnamefont
  {Wang}}\ and\ \bibinfo {author} {\bibfnamefont {Francisco}\ \bibnamefont
  {P\'erez-Bernal}},\ }\bibfield  {title} {\enquote {\bibinfo {title}
  {Characterizing the lipkin-meshkov-glick model excited-state quantum phase
  transition using dynamical and statistical properties of the diagonal
  entropy},}\ }\href {\doibase 10.1103/PhysRevE.103.032109} {\bibfield
  {journal} {\bibinfo  {journal} {Phys. Rev. E}\ }\textbf {\bibinfo {volume}
  {103}},\ \bibinfo {pages} {032109} (\bibinfo {year} {2021})}\BibitemShut
  {NoStop}%
\bibitem [{\citenamefont {Aubry}\ and\ \citenamefont
  {Andr{\'{e}}}(1980)}]{Aubry1980}%
  \BibitemOpen
  \bibfield  {author} {\bibinfo {author} {\bibfnamefont {Serge}\ \bibnamefont
  {Aubry}}\ and\ \bibinfo {author} {\bibfnamefont {Gilles}\ \bibnamefont
  {Andr{\'{e}}}},\ }\bibfield  {title} {\enquote {\bibinfo {title}
  {{Analyticity breaking and Anderson localization in incommensurate
  lattices}},}\ }\href@noop {} {\bibfield  {journal} {\bibinfo  {journal}
  {Proceedings, VIII International Colloquium on Group-Theoretical Methods in
  Physics}\ }\textbf {\bibinfo {volume} {3}} (\bibinfo {year}
  {1980})}\BibitemShut {NoStop}%
\bibitem [{\citenamefont {Harper}(1955)}]{Harper1955}%
  \BibitemOpen
  \bibfield  {author} {\bibinfo {author} {\bibfnamefont {P.~G.}\ \bibnamefont
  {Harper}},\ }\bibfield  {title} {\enquote {\bibinfo {title} {{Single band
  motion of conduction electrons in a uniform magnetic field}},}\ }\href
  {\doibase 10.1088/0370-1298/68/10/304} {\bibfield  {journal} {\bibinfo
  {journal} {Proceedings of the Physical Society. Section A}\ }\textbf
  {\bibinfo {volume} {68}},\ \bibinfo {pages} {874--878} (\bibinfo {year}
  {1955})}\BibitemShut {NoStop}%
\bibitem [{\citenamefont {Dom{\'{\i}}nguez-Castro}\ and\ \citenamefont
  {Paredes}(2019)}]{DomnguezCastro2019}%
  \BibitemOpen
  \bibfield  {author} {\bibinfo {author} {\bibfnamefont {G~A}\ \bibnamefont
  {Dom{\'{\i}}nguez-Castro}}\ and\ \bibinfo {author} {\bibfnamefont
  {R}~\bibnamefont {Paredes}},\ }\bibfield  {title} {\enquote {\bibinfo {title}
  {The aubry{\textendash}andr{\'{e}} model as a hobbyhorse for understanding
  the localization phenomenon},}\ }\href {\doibase 10.1088/1361-6404/ab1670}
  {\bibfield  {journal} {\bibinfo  {journal} {Eur. J. Phys.}\ }\textbf
  {\bibinfo {volume} {40}},\ \bibinfo {pages} {045403} (\bibinfo {year}
  {2019})}\BibitemShut {NoStop}%
\bibitem [{\citenamefont {Lye}\ \emph {et~al.}(2007)\citenamefont {Lye},
  \citenamefont {Fallani}, \citenamefont {Fort}, \citenamefont {Guarrera},
  \citenamefont {Modugno}, \citenamefont {Wiersma},\ and\ \citenamefont
  {Inguscio}}]{Lye2007}%
  \BibitemOpen
  \bibfield  {author} {\bibinfo {author} {\bibfnamefont {J.~E.}\ \bibnamefont
  {Lye}}, \bibinfo {author} {\bibfnamefont {L.}~\bibnamefont {Fallani}},
  \bibinfo {author} {\bibfnamefont {C.}~\bibnamefont {Fort}}, \bibinfo {author}
  {\bibfnamefont {V.}~\bibnamefont {Guarrera}}, \bibinfo {author}
  {\bibfnamefont {M.}~\bibnamefont {Modugno}}, \bibinfo {author} {\bibfnamefont
  {D.~S.}\ \bibnamefont {Wiersma}}, \ and\ \bibinfo {author} {\bibfnamefont
  {M.}~\bibnamefont {Inguscio}},\ }\bibfield  {title} {\enquote {\bibinfo
  {title} {Effect of interactions on the localization of a bose-einstein
  condensate in a quasiperiodic lattice},}\ }\href {\doibase
  10.1103/PhysRevA.75.061603} {\bibfield  {journal} {\bibinfo  {journal} {Phys.
  Rev. A}\ }\textbf {\bibinfo {volume} {75}},\ \bibinfo {pages} {061603}
  (\bibinfo {year} {2007})}\BibitemShut {NoStop}%
\bibitem [{\citenamefont {Tanzi}\ \emph {et~al.}(2013)\citenamefont {Tanzi},
  \citenamefont {Lucioni}, \citenamefont {Chaudhuri}, \citenamefont {Gori},
  \citenamefont {Kumar}, \citenamefont {D'Errico}, \citenamefont {Inguscio},\
  and\ \citenamefont {Modugno}}]{Tanzi2013}%
  \BibitemOpen
  \bibfield  {author} {\bibinfo {author} {\bibfnamefont {Luca}\ \bibnamefont
  {Tanzi}}, \bibinfo {author} {\bibfnamefont {Eleonora}\ \bibnamefont
  {Lucioni}}, \bibinfo {author} {\bibfnamefont {Saptarishi}\ \bibnamefont
  {Chaudhuri}}, \bibinfo {author} {\bibfnamefont {Lorenzo}\ \bibnamefont
  {Gori}}, \bibinfo {author} {\bibfnamefont {Avinash}\ \bibnamefont {Kumar}},
  \bibinfo {author} {\bibfnamefont {Chiara}\ \bibnamefont {D'Errico}}, \bibinfo
  {author} {\bibfnamefont {Massimo}\ \bibnamefont {Inguscio}}, \ and\ \bibinfo
  {author} {\bibfnamefont {Giovanni}\ \bibnamefont {Modugno}},\ }\bibfield
  {title} {\enquote {\bibinfo {title} {Transport of a bose gas in 1d disordered
  lattices at the fluid-insulator transition},}\ }\href {\doibase
  10.1103/PhysRevLett.111.115301} {\bibfield  {journal} {\bibinfo  {journal}
  {Phys. Rev. Lett.}\ }\textbf {\bibinfo {volume} {111}},\ \bibinfo {pages}
  {115301} (\bibinfo {year} {2013})}\BibitemShut {NoStop}%
\bibitem [{Note2()}]{Note2}%
  \BibitemOpen
  \bibinfo {note} {This ensures that while the periodic boundary conditions are
  fulfilled, the potential is aperiodic within the lattice as in experimental
  realizations in optical lattices.}\BibitemShut {Stop}%
\bibitem [{\citenamefont {Styliaris}\ \emph {et~al.}(2019)\citenamefont
  {Styliaris}, \citenamefont {Anand}, \citenamefont {Campos~Venuti},\ and\
  \citenamefont {Zanardi}}]{Styliaris2019}%
  \BibitemOpen
  \bibfield  {author} {\bibinfo {author} {\bibfnamefont {Georgios}\
  \bibnamefont {Styliaris}}, \bibinfo {author} {\bibfnamefont {Namit}\
  \bibnamefont {Anand}}, \bibinfo {author} {\bibfnamefont {Lorenzo}\
  \bibnamefont {Campos~Venuti}}, \ and\ \bibinfo {author} {\bibfnamefont
  {Paolo}\ \bibnamefont {Zanardi}},\ }\bibfield  {title} {\enquote {\bibinfo
  {title} {Quantum coherence and the localization transition},}\ }\href
  {\doibase 10.1103/PhysRevB.100.224204} {\bibfield  {journal} {\bibinfo
  {journal} {Phys. Rev. B}\ }\textbf {\bibinfo {volume} {100}},\ \bibinfo
  {pages} {224204} (\bibinfo {year} {2019})}\BibitemShut {NoStop}%
\bibitem [{\citenamefont {Dooley}\ and\ \citenamefont
  {Kells}(2020)}]{Dooley2020}%
  \BibitemOpen
  \bibfield  {author} {\bibinfo {author} {\bibfnamefont {Shane}\ \bibnamefont
  {Dooley}}\ and\ \bibinfo {author} {\bibfnamefont {Graham}\ \bibnamefont
  {Kells}},\ }\bibfield  {title} {\enquote {\bibinfo {title} {Enhancing the
  effect of quantum many-body scars on dynamics by minimizing the effective
  dimension},}\ }\href {\doibase 10.1103/PhysRevB.102.195114} {\bibfield
  {journal} {\bibinfo  {journal} {Phys. Rev. B}\ }\textbf {\bibinfo {volume}
  {102}},\ \bibinfo {pages} {195114} (\bibinfo {year} {2020})}\BibitemShut
  {NoStop}%
\bibitem [{\citenamefont {Varizi}\ \emph {et~al.}(2021)\citenamefont {Varizi},
  \citenamefont {Cipolla}, \citenamefont {Perarnau-Llobet}, \citenamefont
  {Drumond},\ and\ \citenamefont {Landi}}]{Varizi2021}%
  \BibitemOpen
  \bibfield  {author} {\bibinfo {author} {\bibfnamefont {Adalberto~D}\
  \bibnamefont {Varizi}}, \bibinfo {author} {\bibfnamefont {Mariana~A}\
  \bibnamefont {Cipolla}}, \bibinfo {author} {\bibfnamefont {Mart{\'{\i}}}\
  \bibnamefont {Perarnau-Llobet}}, \bibinfo {author} {\bibfnamefont
  {Raphael~C}\ \bibnamefont {Drumond}}, \ and\ \bibinfo {author} {\bibfnamefont
  {Gabriel~T}\ \bibnamefont {Landi}},\ }\bibfield  {title} {\enquote {\bibinfo
  {title} {Contributions from populations and coherences in non-equilibrium
  entropy production},}\ }\href {\doibase 10.1088/1367-2630/abfe20} {\bibfield
  {journal} {\bibinfo  {journal} {New J. Phys.}\ }\textbf {\bibinfo {volume}
  {23}},\ \bibinfo {pages} {063027} (\bibinfo {year} {2021})}\BibitemShut
  {NoStop}%
\bibitem [{\citenamefont {Scandi}\ \emph {et~al.}(2020)\citenamefont {Scandi},
  \citenamefont {Miller}, \citenamefont {Anders},\ and\ \citenamefont
  {Perarnau-Llobet}}]{Scandi2019}%
  \BibitemOpen
  \bibfield  {author} {\bibinfo {author} {\bibfnamefont {Matteo}\ \bibnamefont
  {Scandi}}, \bibinfo {author} {\bibfnamefont {Harry J.~D.}\ \bibnamefont
  {Miller}}, \bibinfo {author} {\bibfnamefont {Janet}\ \bibnamefont {Anders}},
  \ and\ \bibinfo {author} {\bibfnamefont {Martí}\ \bibnamefont
  {Perarnau-Llobet}},\ }\bibfield  {title} {\enquote {\bibinfo {title}
  {{Quantum work statistics close to equilibrium}},}\ }\href {\doibase
  10.1103/PhysRevResearch.2.023377} {\bibfield  {journal} {\bibinfo  {journal}
  {Phys. Rev. Research}\ }\textbf {\bibinfo {volume} {2}},\ \bibinfo {pages}
  {023377} (\bibinfo {year} {2020})}\BibitemShut {NoStop}%
\bibitem [{\citenamefont {D'Errico}\ \emph {et~al.}(2013)\citenamefont
  {D'Errico}, \citenamefont {Moratti}, \citenamefont {Lucioni}, \citenamefont
  {Tanzi}, \citenamefont {Deissler}, \citenamefont {Inguscio}, \citenamefont
  {Modugno}, \citenamefont {Plenio},\ and\ \citenamefont
  {Caruso}}]{d2013quantum}%
  \BibitemOpen
  \bibfield  {author} {\bibinfo {author} {\bibfnamefont {Chiara}\ \bibnamefont
  {D'Errico}}, \bibinfo {author} {\bibfnamefont {M}~\bibnamefont {Moratti}},
  \bibinfo {author} {\bibfnamefont {E}~\bibnamefont {Lucioni}}, \bibinfo
  {author} {\bibfnamefont {L}~\bibnamefont {Tanzi}}, \bibinfo {author}
  {\bibfnamefont {B}~\bibnamefont {Deissler}}, \bibinfo {author} {\bibfnamefont
  {M}~\bibnamefont {Inguscio}}, \bibinfo {author} {\bibfnamefont
  {G}~\bibnamefont {Modugno}}, \bibinfo {author} {\bibfnamefont {Martin~B}\
  \bibnamefont {Plenio}}, \ and\ \bibinfo {author} {\bibfnamefont
  {F}~\bibnamefont {Caruso}},\ }\bibfield  {title} {\enquote {\bibinfo {title}
  {Quantum diffusion with disorder, noise and interaction},}\ }\href
  {https://doi.org/10.1088/1367-2630/15/4/045007} {\bibfield  {journal}
  {\bibinfo  {journal} {New J. Phys.}\ }\textbf {\bibinfo {volume} {15}},\
  \bibinfo {pages} {045007} (\bibinfo {year} {2013})}\BibitemShut {NoStop}%
\end{thebibliography}%

\pagebreak
\widetext
 
 \newpage 
\begin{center}
\vskip0.5cm
{\Large Supplemental Material}
\end{center}
\vskip0.4cm

\setcounter{section}{0}
\setcounter{equation}{0}
\setcounter{figure}{0}
\setcounter{table}{0}
\setcounter{page}{1}
\renewcommand{\theequation}{S\arabic{equation}}
\renewcommand{\thefigure}{S\arabic{figure}}

\section{Bounds for $H_W$ in terms of $H_{\rm u}$ [Eq.~(\ref{bounds_W_u})]}
In terms of the sets $\Gamma_W=\{(n,m) : E_m^f - E_n^i=W\}$ we can rewrite  Eqs.~\eqref{HW} and~\eqref{Hu} as 
\begin{align}
    H_W &= -\sum_W \sum_{(n,m)\in \Gamma_W} p_n p_{m|n} \ln P(W),
    \\[0.2cm]
    H_{\rm u} &= - \sum_W \sum_{(n,m)\in \Gamma_W} p_n p_{m|n}\ln{p_n p_{m|n}}.
\end{align}
Their difference is then
\begin{align}
    H_{\rm u}-H_{\rm w} &=-\sum_W\sum_{(n,m)\in\Gamma_W}p_n p_{m|n}\ln\left(\frac{p_n p_{m|n}}{P(W)}\right)\label{eq::Ent_dif}\geq 0,
\end{align}
where the inequality follows from the fact that $\frac{p_n p_{m|n}}{P(W)}\leq 1$.
This proves the right-most inequality in Eq.~\eqref{bounds_W_u}. 

To prove the other inequality, we also use Eq.~\eqref{eq::Ent_dif}. Each $\Gamma_W$ on the right hand side is maximal when all of the $p_n p_{m|n}$ in $\Gamma_W$ are equal. So if there are $|\Gamma_W|$ elements in $\Gamma_W$, we get 
\begin{align}
    -\sum_{(n,m)\in\Gamma_W}p_n p_{m|n}\ln\left(\frac{p_n p_{m|n}}{P(W)}\right) &\leq-\sum_{(n,m)\in\Gamma_W}p_n p_{m|n}\ln\left(\frac{1}{|\Gamma_W|}\right)\nonumber =P(W)\ln|\Gamma_W|.
\end{align}
Hence 
\begin{equation}
    H_{\rm u}-H_{\rm w} \leq \sum_W P(W)\ln|\Gamma_W| \leq \ln\gamma_{\rm max}.
\end{equation}

\section{Minimum and maximum work values for the AAH model}
The spectrum of $\mathcal{H}_{\rm AAH}(\Delta)/\hbar$ is bounded between $[-2J-\Delta, 2J+\Delta]$~\cite{d2013quantum}. This bound is rather loose, however. To obtain a tight bound, we note that the spectrum of $\mathcal{H}_{\rm AAH}(\Delta)/\hbar$ lies between  $[-2J-f(\Delta),2J+f(\Delta)]$, where $f(\Delta)$ is a function only of $\Delta$, which can be determined numerically. 
We found that, for $\Delta \lessapprox 4J$, it is very precisely described by 
\begin{equation}
    f(\Delta) = 0.146939 \Delta^2. 
\end{equation}
Focusing on an initial ground state, we then have the following:
\begin{itemize}
    \item $\Delta\to0$: In this case the ground-state has $E_i = -2J-f(\Delta)$, while the final energy (at $\Delta=0$) will range between $E_f \in [-2J,2J]$. 
    Whence, the work $W = E_f - E_i$ must range between
    \begin{align}
        \min(W) &= f(\Delta),\\[0.2cm]
        \max(W) &= 4J + f(\Delta).
    \end{align}
    \item $0\to\Delta$: The ground-state has $E_i=-2J$, while $E_f \in [-2J-f(\Delta), 2J+f(\Delta)]$. Whence, 
    \begin{align*}
         \min(W) &= -f(\Delta),\\[0.2cm]
        \max(W) &= 4J + f(\Delta).
    \end{align*}
\end{itemize}
In both cases, the maximum work is therefore the same. However, in $\Delta\to0$ the minimum work is positive, while in $0\to\Delta$ it is negative. 

\section{Average work for an initial ground state of the AAH model}
The average work can be calculated using the following formula
\begin{equation}
    \langle W\rangle = \tr\left[\left(\mathcal{H}_f-\mathcal{H}_i\right)\ketbra{\psi_0}{\psi_0}\right],
\end{equation}
where $\ket{\psi_0}$ is the ground state of the initial Hamiltonian. Using Eq. \eqref{H_AAH} we can rewrite this as
\begin{equation}
    \langle W\rangle = \Delta \sum_{k=1}^{N}\cos(2\pi\gamma k + \eta) \left|\braket{k}{\psi_0}\right|^2.
\end{equation}
It is clear from this equation that if $\ket{\psi_0}$ is independent of $\Delta$ then the work will scale linearly with $\Delta$, this explains why we only see evidence of the localisation transition when $\mathcal{H}_i$ changes with $\Delta$. When $\mathcal{H}_i = \mathcal{H}_{\rm AAH}(0)$ we can additionally prove that $\langle W\rangle = 0$. This follows from the fact that the ground state is completely delocalised: $\ket{\psi_0} = \frac{1}{\sqrt{N}}\sum_{j=1}^N\ket{j}$. Substituting this in we get
\begin{eqnarray}
    \langle W\rangle 
    &=& \frac{\Delta}{N}\sum_{k=1}^{N}\cos[2\pi\gamma k + \eta]\\
    &=& \frac{\Delta}{N} \mathcal{R}\left[\sum_{k=1}^{N}e^{2\pi i\gamma k + i\eta}\right]\\
    &=& \frac{\Delta}{N} \mathcal{R}\left[e^{i\eta}\frac{e^{i 2\pi\gamma N}-1}{1-e^{-i 2\pi\gamma}}\right] = 0.
\end{eqnarray}
The last equation comes from the fact that we use a rational approximation for the golden ratio given by the ratio of Fibonacci numbers, $\gamma = F_{n-1}/F_n$, and $N=F_n$. In the second line we used Euler's formula and in the third line we used the geometric sum formula.

\end{document}